\documentclass[12pt,letterpaper]{article}
\usepackage[top=0.85in,left=0.85in,footskip=0.75in,marginparwidth=1in]{geometry}
\usepackage{amsmath}
\usepackage{upgreek}
\usepackage{xcolor}
\usepackage{graphicx} 
\usepackage{epstopdf}

\usepackage[utf8]{inputenc}

\usepackage{cite}

\usepackage{nameref,hyperref}

\usepackage{microtype}
\DisableLigatures[f]{encoding = *, family = * }

\raggedright
\usepackage{changepage}

\usepackage[aboveskip=1pt,labelfont=bf,labelsep=period,singlelinecheck=off]{caption}

\makeatletter
\renewcommand{\@biblabel}[1]{\quad#1.}
\makeatother

\usepackage{lastpage,fancyhdr,graphicx}
\usepackage{epstopdf}
\fancyheadoffset[L]{2.25in}
\fancyfootoffset[L]{2.25in}

\usepackage{color}

\definecolor{Gray}{gray}{.25}

\usepackage{graphicx}

\usepackage{sidecap}

\usepackage{wrapfig,cite}
\usepackage[pscoord]{eso-pic}
\usepackage[fulladjust]{marginnote}
\reversemarginpar

\usepackage{ragged2e}
\justifying\let\raggedright\justifying

\begin{document}

\vspace*{0.35in}

\begin{flushleft}
{\Large
\textbf\newline{Slow light, double directional rainbow trapping and releasing at terahertz frequencies}
}
\newline

Jie Xu,\textsuperscript{1,2,3,8} Qian Shen\textsuperscript{2,4,8}, Kai Yuan,\textsuperscript{4} Xiaohua Deng,\textsuperscript{2} Yun Shen,\textsuperscript{2,5} Hang Zhang,\textsuperscript{6} Chiaho Wu,\textsuperscript{6} Sanshui Xiao,\textsuperscript{3,7} and Linfang Shen\textsuperscript{2,6,*}

\bigskip
\bf{1} College of Material Science and Engineering, Nanchang University, Nanchang 330031, China\\
\bf{2} Institute of Space Science and Technology, Nanchang University, Nanchang 330031, China\\
\bf{3} Department of Photonics Engineering, Technical University of Denmark, DK-2800 Kgs. Lyngby, Denmark\\
\bf{4} School of Information Engineer, Nanchang University, Nanchang, 330031, China\\
\bf{5} Department of Physics, Nanchang University, Nanchang 330031, China\\
\bf{6} Department of Applied Physics, Zhejiang University of Technology, Hangzhou 310023, China\\
\bf{7} Center for Nanostructured Graphene, Technical University of Denmark, DK-2800 Kgs. Lyngby, Denmark \\
\bf{8} These authors contributed equally to the work\\
\bigskip
* lfshen@zjut.edu.cn

\end{flushleft}

\section*{Abstract}
Slow light and rainbow trapping attract many attentions in last twenty years, and in most of the previous works, the researchers achieve the slow light and rainbow trapping with complicate configurations or techniques, for example, metamaterial techniques. In this paper, we propose a simple waveguide consisted with semiconductor and dielectric layers, and we find that the slow-light peaks appear in the dispersion curves. Besides, the cutoff frequency with $v_g = 0$ ($v_g$ is the group velocity) in the slow-light peaks is depending on the thicknesses of the semiconductor and dielectric. Then, we design a tapered, horizontally symmetric structure. By using software COMSOL and finite difference time domain method as well, we achieve double directional rainbow trapping and more importantly, releasing the rainbow in such structure. This approach is promising for optical isolator, optical buffer, optical switch and other optical functional devices in miniaturization optical integrated circuit.


\section{Introduction}
Surface magnetoplasmon (SMP) is sustained at the interface of a magnetic-optical (MO) material and a dielectric material, and it will behave the one-way propagating property when applying a dc external magnetic field on the MO material. The one-way SMP is similar to the chiral edge states founded in the quantum-Hall effect \cite{Prang:1}. In 2008, F. D. M. Haldane and S. Raghu showed the possibility of one-way waveguide in photonic crystals (PCs)\cite{Haldane:2}, and in the next year, Wang and his colleagues experimentally observed the one-way SMP in a PC consisted of gyromagnetic yttrium-iron-garnet (YIG) in microwave regime\cite{Wang:3}. Since then, one-way SMP draws more and more attentions\cite{Qiu:4, Shen:5,Jin:6,Liu:7}. 

Manipulating the light always attracts the physicists and slowing light is one of the most interesting and meaningful part because of its potential applications in energy storage\cite{Tsakmakidis:8,Su:9,Su:10}, enhancing nonlinearity\cite{Sol:11,Heebner:12,Chen:13}, quantum optics\cite{Ku:14,Wu:15}. At the very beginning, people study the slow-light by using electromagnetically induced transparency (EIT) which is excellent for fundamental investigations, but is unsuitable for practical applications\cite{Krauss:16,Dutton:17}. In the past decades, lots of researchers are trying to slow light in PCs by introducing defects and the photonic bandgap (PBG)\cite{Baba:18,Krauss:19,Sol:11,Hosseini:20}, and others used the metamaterials\cite{Tsakmakidis:8}. Recently, we proposed a simple tapered metal-dielectric-semiconductor-metal (MDSM) structure in terahertz regime\cite{Xu:21}, and we proved that this MDSM structure is capable for slowing light (SMP) and trapping 'rainbow'.

In this paper, we propose a metal-dielectric-semiconductor-dielectric-metal (MDSDM) structure and theoretically study the propagation properties in the MDSDM model. After carefully investigation, we find that the dispersion properties of the SMPs in the MDSDM model is manipulable by changing the thicknesses of the dielectric and the semiconductor layers. Based on our theory, we design a tapered, horizontal symmetric waveguide and by using the software COMSOL and finite difference time domain (FDTD) method, for the first time, we present the double directional slow light, rainbow trapping and releasing in terahertz regime. Our work is promising for optical functional devices.

\section{Physical model and slow light}
\begin{figure}[ht]
\centering\includegraphics[width=4.5 in]{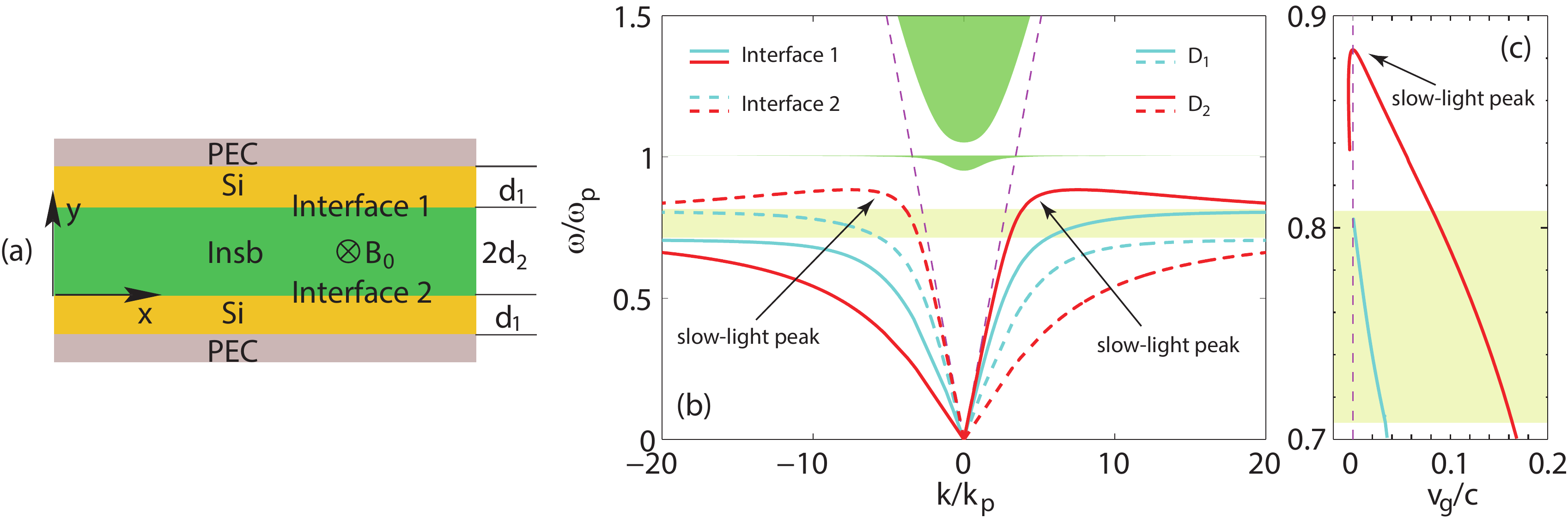}
\caption{ (a) The schematic of the MDSDM structure. (b) The dispersion curves of SMPs. Red lines: $D_1=(0.025\lambda_p,0.025\lambda_p)$; Cyan lines: $D_2=(0.025\lambda_p,0.006\lambda_p)$. The solid and dashed lines are respectively represent the SMPs sustained on interfaces 1 and 2. (c) The group velocity ($v_g$) of the upper branches of SMPs dispersion curves in (b) for $k\geq 0$. The yellow shaded areas are the asymptotic frequencies band and the green shaded areas represent the bulk zones of the semiconductor. The waveguide parameter $D=(d_1,d_2)$. The other parameters are $\varepsilon_r=11.68$, $\varepsilon_\infty=15.6$ and $\omega_c=0.1\omega_p$. }\label{Fig1}
\end{figure}
We first investigate the dispersion relation of the SMPs in the metal-dielectric-semiconductor-dielectric-metal (MDSDM) waveguide shown in Fig. 1(a). We note that, in the terahertz regime, metal can be regarded as perfect electric conductor (PEC)\cite{Shen:22} and only the transverse magnetic (TM) modes can propagate along dielectric-semiconductor interfaces. The (relative) permittivity of the semiconductor (in this paper we assume the semiconductor is Insb and the dielectric is Si) under a dc magnetic field ($B_0$) has the form\cite{Brion:23}
\begin{equation}
\mathop \varepsilon \limits^ \leftrightarrow  = \left[ {\begin{array}{*{20}{c}}
{{\varepsilon _1}}&{-i{\varepsilon _2}}&0\\
{i{\varepsilon _2}}&{{\varepsilon _1}}&0\\
0&0&{{\varepsilon _3}}
\end{array}} \right], \tag{1a}
\end{equation}

\begin{equation}
    \varepsilon_{1}  = \varepsilon _\infty  \left( {1 - \frac{{\omega
    _p^2 }}{{\omega ^2  - \omega _c^2 }}} \right),\varepsilon_{2}  = \varepsilon _\infty  \frac{{\omega_c\omega _p^2
    }}{{\omega (\omega ^2  - \omega _c^2 )}},
    \varepsilon_{3}  = \varepsilon _\infty  \left( {1 - \frac{{\omega
    _p^2 }}{{\omega ^2 }}} \right), \tag{1b}
\end{equation}
\begin{equation}
    \varepsilon_1=\varepsilon_\infty\left(1-\frac{\left(\omega+iv\right)\omega_p^2}{\omega\left[\left(\omega+iv\right)^2-\omega_c^2\right]}\right),\;\varepsilon_2=\varepsilon_\infty\frac{\omega_c\omega_p^2}{\omega\left[\left(\omega+iv\right)^2-\omega_c^2\right]},
    \varepsilon_3=\varepsilon_\infty\left(1-\frac{\omega_p^2}{\omega\left(\omega+iv\right)}\right),\tag{1c}
\end{equation}
where $\omega$, $\omega_p$, $\omega_c=eB_0/m^*$, $e$, $m^*$ and $\varepsilon_\infty$  are the angular frequency, the plasma frequency, the electron cyclotron frequency, the charge of a electron, the effective mass of a electron and the high-frequency (relative) permittivity of the semiconductor, respectively. Eq. (1c) is the permittivity of the semiconductor in loss condition and $\nu$ is the electron scattering frequency. We analysis the dispersion relation in lossless condition and in the simulations below, we will show that the propagation properties in loss condition fit well with the lossless condition. Based on the Maxwell's equations and the boundary conditions, one can easily obtain the dispersion relation of the SMPs in the MDSDM model, and it can be written as
\begin{equation}
    \frac{2\varepsilon_v\alpha_1}{\varepsilon_r}\tanh(\alpha_1d_1)+\frac A\alpha\tanh(2\alpha d_2)=0, \tag{2a}
\end{equation}

\begin{equation}
    A=(\frac{2\varepsilon_v\alpha_1}{\varepsilon_r}\tanh{(\alpha_1d_1))}^2-(\frac{\varepsilon_2}{\varepsilon_1}k)^2+\alpha^2\tag{2b}
\end{equation}
where $\varepsilon_v = \varepsilon_1 - \frac{\varepsilon_2^2}{\varepsilon_1}$, $\varepsilon_r$, $\alpha=\sqrt{k^2-\varepsilon_v k_0^2}$, $\alpha_1=\sqrt{k^2-\varepsilon_r k_0^2}$ and $d_1$ ($d_2$) are the Voigt permittivity, permittivity of the dielectric, the attenuation coefficient of the SMPs in the semiconductor layer, the attenuation coefficient of the SMPs in the dielectric layer and the thickness of the dielectric (semiconductor), respectively. Eq. (2) has no $k$ term, which is because there are two opposed symmetric interfaces which we name them interfaces 1 and 2 (see Fig. 1(a)) and both of them can sustain the SMPs. For $k \to \pm \infty$, we calculate the asymptotic frequencies $\omega_{sp}^{(1)}$ and $\omega_{sp}^{(2)}$ from Eq. (2), and they have the form
\begin{equation}
\omega _{sp}^{(1)}  = \frac{1}{2}\left( {\sqrt {\omega _c^2 + 4\frac{{{\varepsilon _\infty }}}{{{\varepsilon _\infty } + {\varepsilon _r}}}\omega _p^2}  + {\omega _c}} \right),\tag{3a}
\end{equation}
\begin{equation}
\omega _{sp}^{(2)}  = \frac{1}{2}\left( {\sqrt {\omega _c^2 + 4\frac{{{\varepsilon _\infty }}}{{{\varepsilon _\infty } + {\varepsilon _r}}}\omega _p^2}  - {\omega _c}} \right).\tag{3b}
\end{equation}

\begin{figure}[pt]
\centering\includegraphics[width=4.5 in]{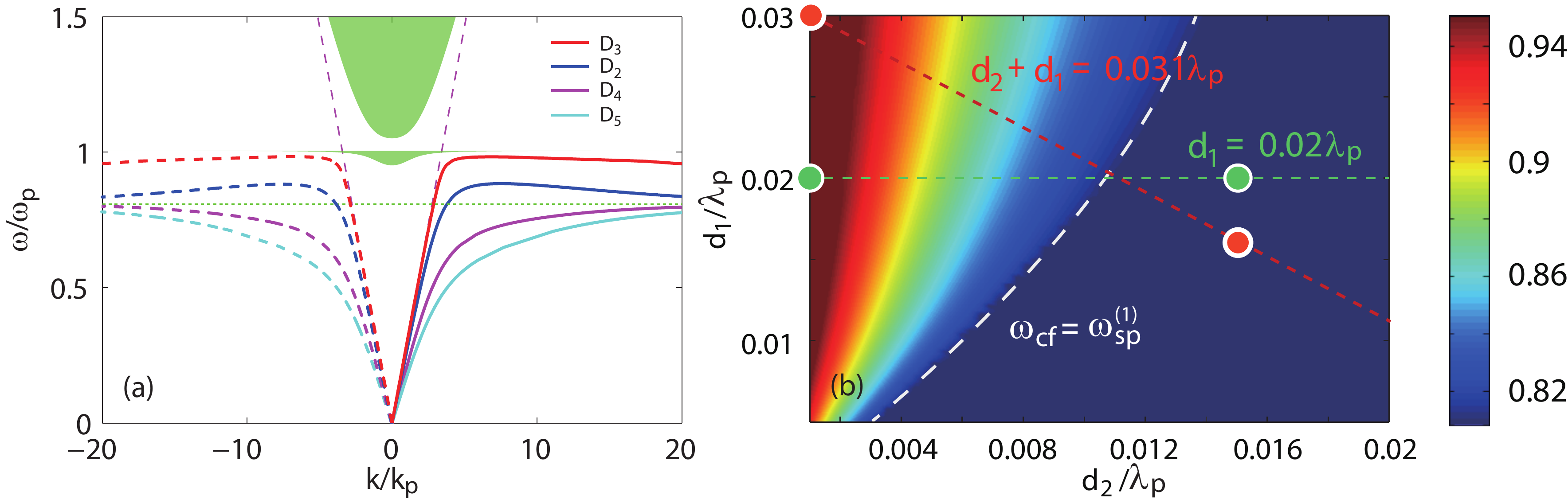}
\caption{ (a) The upper SMP branch as $D_3=(0.03\lambda_p,0.001\lambda_p)$, $D_2=(0.025\lambda_p,0.006\lambda_p)$, $D_4=(0.015\lambda_p,0.016\lambda_p)$ and $D_5=(0.01\lambda_p,0.021\lambda_p)$. The dotted and dashed lines are respectively represent $\omega=\omega_{sp}^{(1)}$ and the light line of the glass. (b) The cutoff frequencies $\omega_{cf}$ as functions of $D=(d_1,d_2)$. The white dashed line represent the interface between $\omega_{cf}=\omega_{sp}^{(1)}$ and $\omega_{cf}>\omega_{sp}^{(1)}$.}\label{Fig2}
\end{figure}

For simplify, we introduce a waveguide parameter $D=(d_1,d_2)$. Fig. 1(b) shows the dispersion curves of SMPs as $D_1=(0.025\lambda_p,0.025\lambda_p)$ (the cyan lines) ($\lambda_p=2\pi c/\omega_p$, c is the light speed in vacuum) and $D_2=(0.025\lambda_p,0.006\lambda_p)$ (the red lines). The shaded yellow area represent the asymptotic frequencies (AF) band and for the $D_1$ case, there are no SMPs over the AF band. In contrast, in the $D_2$ case, the thickness of the semiconductor is thinner than the one in $D_1$ case and we can see clear slow-light peaks appear in the upper branches of the SMPs' dispersion curves. The slow-light peak appears at finite $k$ and more interestingly, there is no SMPs modes in the nearby region upon the peak, and this is quite different with our previous work\cite{Xu:21}, in which we reported upon slow-light peak, there are still SMPs. In Fig. 1(c), we present the group velocity ($v_g$) of the upper branches of the SMPs' dispersion curves of Fig. 1(b) (because of the symmetric properties, here, we just concern $k \geq 0$ area) and one can easily find that in the $D_1$ case, $v_g$ is always passitive and $v_g \to 0$ as $k \to \infty$, and in the $D_2$ case, $v_g$ can reach zero around the slow-light peak.

To further investigate the relation between the slow-light peak and waveguide parameter D, we plot the upper branches of the dispersion curves of SMPs in Fig. 2(a) as $d_1+d_2=0.031\lambda_p$, and $D_3=(0.03\lambda_p,0.001\lambda_p)$, $D_2=(0.025\lambda_p,0.006\lambda_p)$, $D_4=(0.015\lambda_p,0.016\lambda_p)$ and $D_5=(0.01\lambda_p,0.021\lambda_p)$. In Fig. 2(a), the slow-light peaks appear in $D_2$ and $D_3$ cases and the cutoff frequencies $\omega_{cf}$ ($v_g=0$) of the peaks in this two cases are different, more specifically, we deduce that the smaller $d_2$ or the larger $d_1$ the larger $\omega_{cf}$.
In Fig. 2(b), we further demonstrate the relation between $\omega_{cf}$ and $D=(d_1,d_2)$ in ultra thin waveguide condition ($d_1, d_2 \sim 10^{-2}\lambda_p$), and we highlight the edge (the white dashed line) of $\omega_{cf}=\omega_{sp}^{(1)}$ region, and the left area of the white line represent the cases with $\omega_{cf}>\omega_{sp}^{(1)}$. As we aspect, in the $\omega_{cf}>\omega_{sp}^{(1)}$ area, both $d_1$ and $d_2$ can affect $\omega_{cf}$ and the larger $d_1$ or smaller $d_2$ the larger $\omega_{cf}$. 


\section{Double directional rainbow trapping and releasing}

Fig. 2 shows that the cutoff frequencies will change with the waveguide parameter D, and this result present a possible way to achieve rainbow trapping. Fig. 3 is our designed tapered waveguides which is consisted of two Si layers and in the middle of the Si layers is Insb layer with an external magnetic field $B_0$. The length of the straight part is $L$ and two tapered parts have the same length $L_1$. Edges 1 and 2 have different waveguide parameter D, i.e. $D_1 = (0.03\lambda_p,0.001\lambda_p)$ and $D_2 = (0.016\lambda_p,0.015\lambda_p)$. Moreover, the red dashed line in Fig. 2(b) represents $d_1+d_2=0.031\lambda_p$ and the upper left red point represents case $D_1$ (the Edge 1) and another red point shows case $D_2$ (the Edge 2). Besides, derived from Fig. 2(b) we know that $\omega_{cf}$ are $0.98\omega_p$ (the largest value on the red line) for Edge 1 and $0.81\omega_p$ ($=\omega_{\rm sp}^{(1)}$) for Edge 2. Thus, the cutoff frequencies in this structure must satisfy $0.81\omega_p \leq \omega_{cf} \leq 0.98\omega_p$. 

We perform the full wave simulations in loss condition with $v=0.001\omega_p$ by using commercial software COMSOL and Fig. 4 shows the electric field distributions in four different operating frequencies, i.e., (a) $\omega=0.82\omega_p$, (b) $\omega=0.85\omega_p$, (c) $\omega=0.88\omega_p$ and (d)  $\omega=0.94\omega_p$. As we expected, when we put a magnetic current source in the middle of waveguide, clear double directional rainbow trapping is achieved in the designed waveguide and the electromagnetic (EM) mode with higher working frequency ($\omega>\omega_{sp}^{(1)}$) will be trapped at thinner Insb position. As shown in Fig .4, one can believe in that this waveguide is capable to trap the EM modes when the frequencies are lied in the frequency band limited by the cutoff frequencies of Edges 1 and 2. Here we emphasis that, even the similar double directional slow light (EM trapping) can be achieved in a all dielectric waveguide, double directional rainbow trapping has never been achieved or reported.

\begin{figure}[pt]
\centering\includegraphics[width=4.5 in]{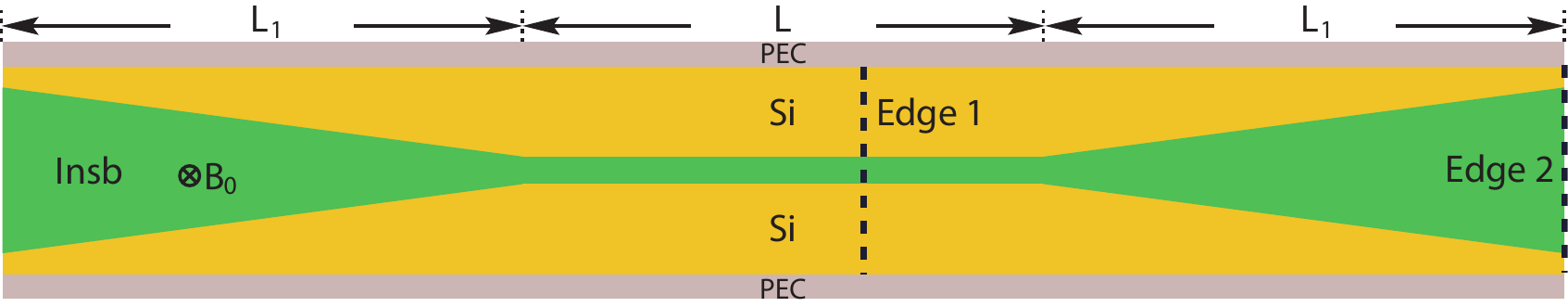}
\caption{ The schematic of the double directional rainbow trapping structure. The tapered part and straight part have the same length, i.e., $L_1=L=100\mu$m. The parameter D of Edge 1 and Edge 2 are $D_1=(0.03\lambda_p,0.001\lambda_p)$ (the left red point shown in Fig. 2(b)) and $D_2=(0.016\lambda_p,0.015\lambda_p)$ (the right red point shown in Fig. 2(b)), respectively.}\label{Fig3}
\end{figure}

\begin{figure}[ht]
\centering\includegraphics[width=4.5 in]{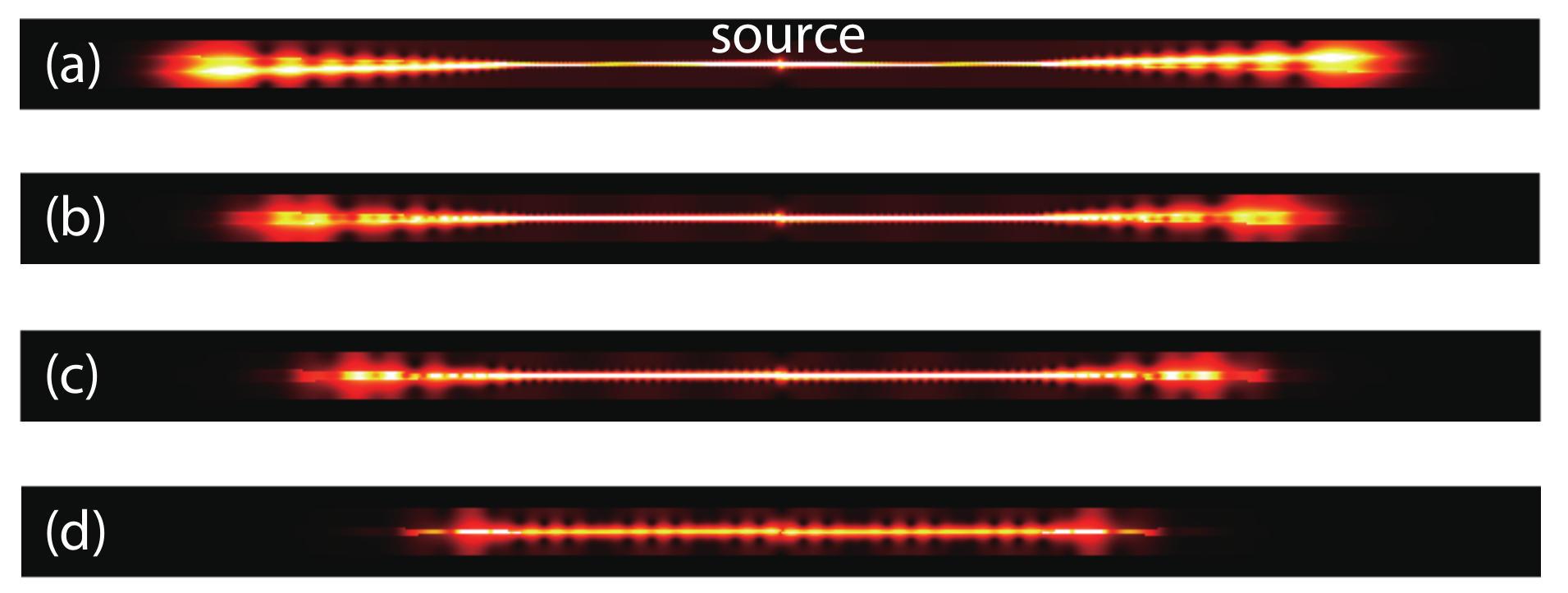}
\caption{ The electric field distribution of double directional rainbow trapping as (a) $\omega=0.82\omega_p$, (b) $\omega=0.85\omega_p$, (c)  $\omega=0.88\omega_p$ and (d)  $\omega=0.94\omega_p$. $\omega$ is the working frequency of the magnetic current source. $\omega_c=0.1\omega_p$ and $\nu = 0.001\omega_p$.}\label{Fig4}
\end{figure}

\begin{figure}[ht]
\centering\includegraphics[width=5.5 in]{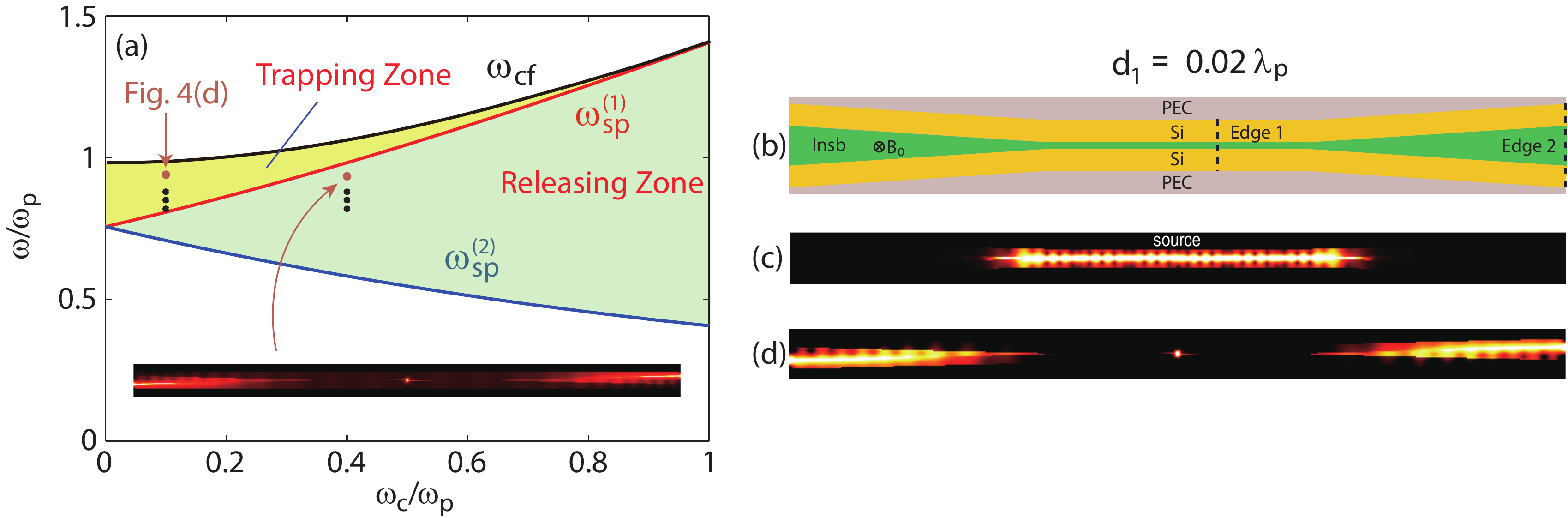}
\caption{ (a) EM energy trapping and releasing theory. The red, blue and black lines are respectively the asymptotic frequencies $\omega_{sp}^{(1)}$, $\omega_{sp}^{(2)}$ and the cutoff frequency $\omega_{cf}$. The inset shows the electric field distribution as $\omega_c=0.4\omega_p$ with $\omega=0.94\omega_p$. The waveguide parameter $D=(0.03\lambda_p,0.001\lambda_p)$. (b) Another double directional rainbow trapping and releasing waveguide. The thickness of the Si layers are constantly being $d_1=0.02\lambda_p$ and as shown in Fig. 2(b), the green points represent the parameters D of Edges 1 and 2 of this waveguide, more specifically, $D_1=(0.02\lambda_p,0.001\lambda_p)$ and $D_2=(0.02\lambda_p,0.015\lambda_p)$. The electric fields distribution in the simulations as $\omega=0.94\omega_p$ with (c) $\omega_c=0.1\omega_p$ and (d) $\omega_c=0.4\omega_p$.}\label{Fig5}
\end{figure}

\begin{figure}[ht]
\centering\includegraphics[width=4.5 in]{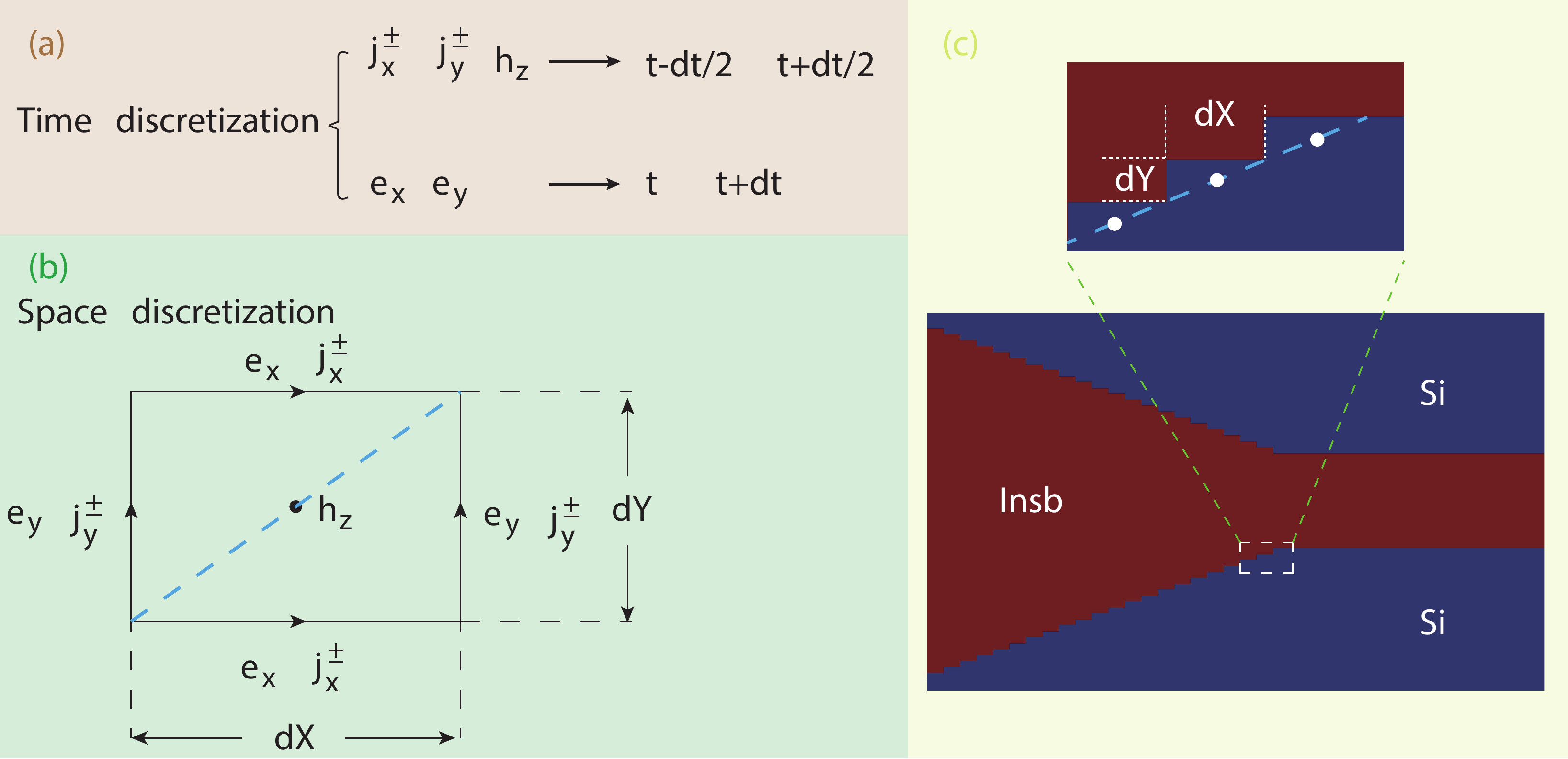}
\caption{ FDTD method in tapered configuration. The (a) time and (b) space discretizations based on Eqs. (4) and (5). (c) A part of the space discretization of the whole waveguide shown in Fig. 3. The blue dashed line shows the Insb-Si interface in real space.}\label{Fig6}
\end{figure}
\begin{figure}[ht]
\centering\includegraphics[width=4.5 in]{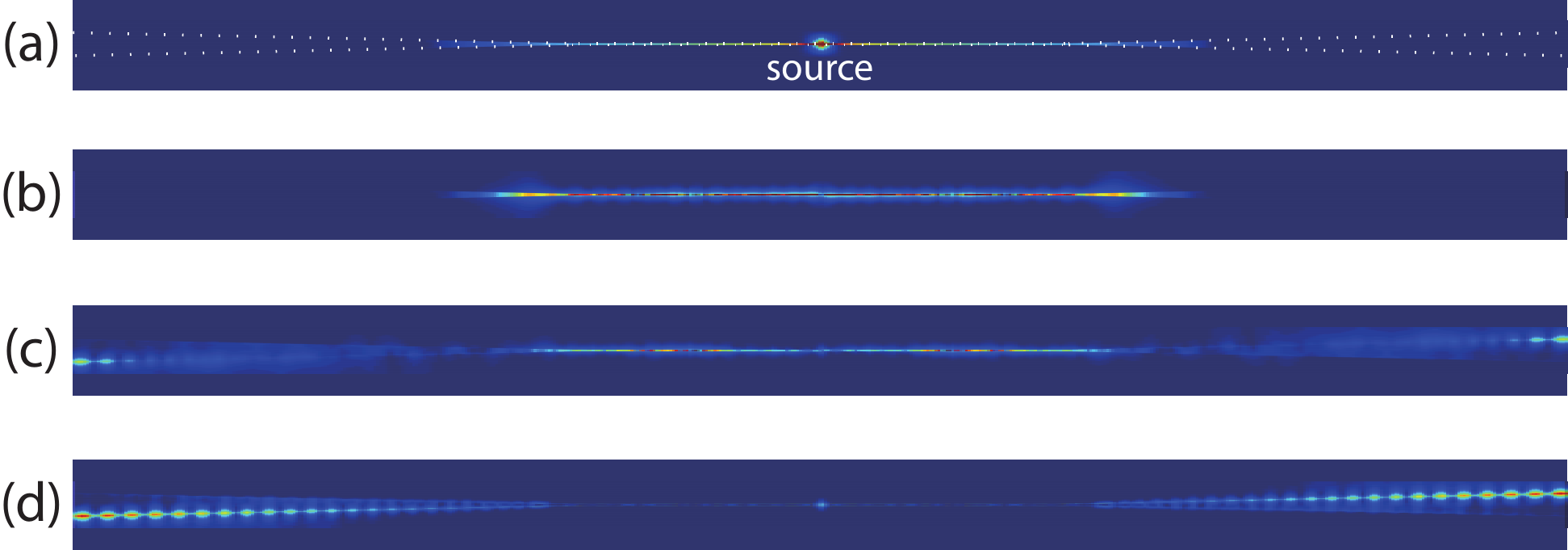}
\caption{ The electric field distribution in FDTD simulations when (a) $t=5T_p$ ($T_p=2\pi/\omega_p$), (b) $t=200T_p$, (c) $t=300T_p$ and (d) $t=1000T_p$. We set $\omega_c=0.1\omega_p$ ($t \leq 200T_p$) and $\omega_c=0.4\omega_p$ ($t>200T_p$). The other parameters are the same with Fig. 4(d).}\label{Fig7}
\end{figure}

Once we trapped the EM modes, the following question is how to release them. Since the optical properties of the MO materials can be engineered by changing the external magnetic field, in Fig. 5(a), as $D=(0.03\lambda_p,0.001\lambda_p)$, we plot the asymptotic frequencies and cutoff frequency $\omega_{cf}$ as functions of $\omega_c$ which has linear relation with the external magnetic field. The green and yellow shaded zones represent $\omega_{\rm sp}^{(1)}<\omega<\omega_{cf}$ and $\omega_{\rm sp}^{(2)}<\omega<\omega_{\rm sp}^{(1)}$, and we name them the trapping zone and releasing zone (the AF region), respectively. The left four points corresponding to the four working frequencies in Fig. 4 and they all fall in the trapping zone when $\omega_c=0.1\omega_p$. If we enlarge the external magnetic field and make $\omega_c=0.4\omega_p$ (the right four points), all of the four frequencies fall in the AF region, and the SMPs lies in the AF region can always propagate along the Insb-Si interfaces, therefore, $|v_g|>0$ which implies that the SMPs can not be trapped in the waveguide. As an example, we perform the simulation as shown the inset in Fig. 5(a) as $\omega_c=0.4\omega_p$ and $\omega=0.94\omega_p$, and it is quite clear that the SMPs excited by the source propagate to the end surface but rather be trapped. Besides, other waveguides based on our theory, e.g. the one shown in Fig. 5(b) in which the thickness of Si layers are constant ($d_1=0.02\lambda_p$, the horizontal green line in Fig. 2(b)) can also achieve the double directional rainbow trapping (Fig. 5(c), $\omega_c=0.1\omega_p$) (here, we just show one of the rainbow trapping) and releasing (Fig. 5(d), $\omega_c=0.4\omega_p$) as $\omega=0.94\omega_p$. 

To clearly show the process of trapping and releasing the EM energy, we use the FDTD method re-perform the simulations. For the semiconductor, combing Eq. (1) and Maxwell's equations, we can obtain
 
 \begin{equation}
    \frac{\partial j_x^\pm}{\partial t}+2\pi(\overline\nu\mp i\overline\omega_c)j_x^\pm=2\pi^2\varepsilon_\infty e_x, \tag{4a}
 \end{equation}
 \begin{equation}
     \frac{\partial j_y^\pm}{\partial t}+2\pi(\overline\nu\mp i\overline\omega_c)j_y^\pm=2\pi^2\varepsilon_\infty e_y, \tag{4b}
 \end{equation}
 \begin{equation}
     \frac{\partial h_z}{\partial Y}=\varepsilon_\infty\frac{\partial e_x}{\partial t}+(j_x^++j_x^-)-i(j_y^+-j_y^-), \tag{4c}
 \end{equation}
 \begin{equation}
     -\frac{\partial h_z}{\partial X}=\varepsilon_\infty\frac{\partial e_y}{\partial t}+(j_y^++j_y^-)+i(j_x^+-j_x^-), \tag{4d}
 \end{equation}
 \begin{equation}
     \frac{\partial e_y}{\partial X}-\frac{\partial e_x}{\partial Y}=-\frac{\partial hz}{\partial t}, \tag{4e}
 \end{equation}
where $\overline \omega_c=\omega_c/\omega_p$, $j_x$ and $j_y$ are the induced current components. ($e_x,e_y,h_z$) are the normalized electric field and magnetic field. For the Si layer, we can easily get the similar equations from Eqs. (4c,4d,4e) by setting $j_x^\pm=0$, $j_y^\pm=0$ and letting $\varepsilon_r$ replace $\varepsilon_\infty$. Then, for the Si layer, we have
\begin{equation}
     \frac{\partial h_z}{\partial Y}=\varepsilon_r\frac{\partial e_x}{\partial t}, \tag{5a}
 \end{equation}
 \begin{equation}
     -\frac{\partial h_z}{\partial X}=\varepsilon_r\frac{\partial e_y}{\partial t}, \tag{5b}
 \end{equation}
 \begin{equation}
     \frac{\partial e_y}{\partial X}-\frac{\partial e_x}{\partial Y}=-\frac{\partial hz}{\partial t}, \tag{5c}
 \end{equation}
Based on Eqs. (4) and (5), we can discrete ($e_x,e_y,h_z,j_x^\pm,j_y^\pm$) in time domain and space domain as shown in Figs. 6(a,b). One of challenges in applying the FDTD method in such tapered structure is how to define the boundary conditions especially on the tapered Insb-YIG interfaces. Here, we design zig-zag interfaces (see the inset in Fig. 6(c)) instead of the original smooth interfaces (the blue dashed lines in Fig. 6) in real space. We note that around the zig-zag interfaces, if we set suitable ratio $dX/dY$ and let the $h_z$ of all the units (the white points ($h_z$) in Fig. 6(c)) that through the original interface lie on the original interface, the boundary condition is the same with the one in real space, i.e. $h_z$ is continuous on the Insb-Si interfaces. Therefore, as shown in Fig. 6(c), we design the space discretization for structure shown in Fig. 3 (here, we only show the straight-tapered joint part of the structure). Based on the Eq. (4), Eq. (5) and the time-space discretizations shown in Fig. 6, we perform the FDTD simulations in Fig. 7. Fig. 7(a), (b), (c) and (d) show the electric field distributions when $t=5T_p$ ($T_p=2\pi/\omega_p$), $t=200T_p$, $t=300T_p$ and $t=1000T_p$, respectively. Note that we set $\omega_c=0.1\omega_p$ for $t<=200T_p$ and as shown in Figs. 7(a,b), the EM energy is truly trapped in the waveguide which are perfectly fit with our previous result (Fig. 4(d)) solved in COMSOL. For $t>200T_p$, we set $\omega_c=0.4\omega_p$ and Fig. 7(c) shows that the trapped EM energy is no longer trapped but gradually released. When $t=1000T_p$ the EM energy is mostly localized on the end surfaces which are set to be PEC in these simulations, which is also the same with the result shown in the inset of Fig. 5(a). This kind of rainbow trapping and releasing structure is promising for designing some functional devices and probably can be used to break the time-bandwidth limit (not considered here). 

\section{Conclusion}
In conclusion, we propose a metal-dielectric-semiconductor (under an external magnetic field $B_0$)-dielectric-metal (MDSDM) configuration. In our analysis, we find that slow-light peaks appear in the thin cases ($d_1 \times d_2<0.001\lambda_p$, $d_1$ and $d_2$ are respectively the thicknesses of the semiconductor and the dielectric) and the cutoff frequencies ($\omega_{cf}$) can change with $D=(d_1,d_2)$. Then, we design a tapered, horizontal symmetric waveguide and in the (COMSOL) simulations, the EM modes are trapped in different locations for four different frequencies, i.e. $\omega=0.82\omega_p$, $0.85\omega_p$, $0.88\omega_p$ and $0.94\omega_p$ as $\omega_c=0.1\omega_p$ ($\omega_c=eB_0/m^*$, where e and $m^*$ are, respectively, the charge and effective mass of a electron). More interestingly, when we enlarge the $B_0$ and let $\omega_c = 0.4\omega_p$, the trapped energy will be released. We also propose a method to use finite difference time domain(FDTD) method in the designed tapered structure and perform the simulations in time domain, and as a result, clear EM energy trapping and releasing are observed. 

\section*{Funding information}
We acknowledge support by National Natural Science Foundation of China (NSFC) (61372005), National Natural Science Foundation of China (NSFC) under a key project (41331070), and Independent Research Fund Denmark (9041-00333B). The Center for Nanostructured Graphene is sponsored by the Danish National Research Foundation (Project No. DNRF103).



\begin{thebibliography}{99}

\bibitem{Prang:1}
R.~E. Prange and S.~M. Girvin, \emph{The Quantum Hall effect} (Springer, 1987).

\bibitem{Haldane:2}
F.~D.~M. Haldane and S.~Raghu, "Possible realization of directional
  optical waveguides in photonic crystals with broken time-reversal symmetry,"
  Phys. Rev. Lett. \textbf{100}(1), 013904
  (2008).

\bibitem{Wang:3}
Z.~Wang, Y.~Chong, J.~D. Joannopoulos, and M.~Solja{\v{c}}i{\'c},
  "Observation of unidirectional backscattering-immune topological
  electromagnetic states," Nature \textbf{461}(7265),
  772 (2009).

\bibitem{Qiu:4}
W. Qiu, Z. Wang, and M.~Solja{\v{c}}i{\'c}, "Broadband circulators based on directional coupling of one-way waveguides," Opt. Express \textbf{19}(22), 22248-22257 (2011).

\bibitem{Shen:5}
L. Shen, Y. You, Z. Wang, and X. Deng, "Backscattering-immune one-way surface magnetoplasmons at terahertz frequencies," Opt. Express \textbf{23}(2), 950--962 (2015).

\bibitem{Jin:6}
D. Jin, L. Lu, Z. Wang, C. Fang, J. D. Joannopoulos, M.~Solja{\v{c}}i{\'c}, L. Fu, and N. X. Fang, "Topological magnetoplasmon," Nat. Commun. \textbf{7}, 13486 (2016).

\bibitem{Liu:7}
K. Liu, A. Torki, and S. He, "One-way surface magnetoplasmon cavity and its application for nonreciprocal devices," Opt. Lett. \textbf{41}(4), 800--803 (2016).

\bibitem{Tsakmakidis:8}
 K. L. Tsakmakidis, A. D. Boardman, and O. Hess, "'Trapped rainbow' storage of light in metamaterials," Nature \textbf{450}, 397--401 (2007).

\bibitem{Su:9}
Y. Su, F. Liu, and Q. Li, "System performance of slow-light buffering and storage in silicon nano-waveguide," Proc. SPIE \textbf{6783}, 67832--67832 (2007).

\bibitem{Su:10}
S. W. Su, Y. H. Chen, S. C. Gou, T. L. Horng, and I. A. Yu, "Dynamics of slow light and light storage in a Doppler-broadened electromagnetically induced-transparency medium: a numerical approach," Phys. Rev. A \textbf{83}, 013827 (2011).

\bibitem{Sol:11}
M.~Solja{\v{c}}i{\'c}, S. G. Johnson, S. Fan, M. Ibanescu, E. Ippen, and J. D. Joannopoulos, "Photonic-crystal slow-light enhancement of nonlinear phase sensitivity," J. Opt. Soc. Am. B \textbf{19}(9), 2052--2059 (2002).

\bibitem{Heebner:12}
J. E. Heebner, R. W. Boyd, and Q. H. Park, "Slow light, induced dispersion, enhanced nonlinearity, and optical solitons in a resonator-array waveguide," Phys. Rev. E-Statistical, Nonlinear Soft Matter. Phys. \textbf{65}(3), 0366191--0366194 (2002).

\bibitem{Chen:13}
Y. Chen and S. Blair, "Nonlinearity enhanced in finite coupled-resonator slow-light waveguides," Opt.
Express \textbf{12}, 3353--3366 (2004).

\bibitem{Ku:14}
P. Ku, F. Sedgwick, C. J. Chang-Hasnain, P. Palinginis, T. Li, H. L. Wang, S. W. Chang, and S. L. Chuang, "Slow light in semiconductor quantum wells," Opt. Lett. 29, 2291--2293 (2004).

\bibitem{Wu:15}
B. Wu, J. F. Hulbert, E. J. Lunt, K. Hurd, A. R. Hawkins, and H. Schmidt, "Slow light on a chip via atomic quantum state control," Nat. Photonics \textbf{4}, 776--779 (2010).

\bibitem{Krauss:16} 
T. F. Krauss, "Why do we need slow light?" Nat. Photonics \textbf{2}(8), 448 (2008).

\bibitem{Dutton:17}
Z. Dutton, M. Budde, C. Slowe, and L. V. Hau, "Observation of quantum shock waves created with ultra-compressed slow light pulses in a Bose-Einstein condensate," Science \textbf{293}(5530), 663--668 (2001).

\bibitem{Baba:18}
T. Baba, "Slow light in photonic crystals," Nat. Photonics \textbf{2}(8), 465 (2008).

\bibitem{Krauss:19}
T. F. Krauss, "Slow light in photonic crystal waveguides," J. Phys. D Appl. Phys. \textbf{40}(9), 2666 (2007).

\bibitem{Hosseini:20}
A. Hosseini, X. Xu, D. N. Kwong, H. Subbaraman, W. Jiang, and R. T. Chen, "On the role of evanescent modes and group index tapering in slow light photonic crystal waveguide coupling efficiency," Appl. Phys. Lett. \textbf{98}(3), 031107 (2011).

\bibitem{Xu:21}
J.~Xu, S.~Xiao, C.~Wu, H.~Zhang, X.~Deng, and L.~Shen, "Broadband
  one-way propagation and rainbow trapping of terahertz radiations,"
  Opt. Express \textbf{27}(8), 10659--10669 (2019).

\bibitem{Shen:22}
L. Shen, J. Xu, Y. You, K. Yuan, and X. Deng, "One-way electromagnetic mode guided by the mechanism of total internal reflection,"  IEEE Photonics Technol. Lett. \textbf{30}(2), 133--136 (2018).

\bibitem{Brion:23}
 J. J. Brion, R. F. Wallis, A. Hartstein, and E. Burstein, "Theory of surface magnetoplasmons in semiconductors,"
Phys. Rev. Lett. \textbf{28}(22), 1455--1458 (1972).


\end{thebibliography}

\end{document}